\documentclass{article}
\usepackage{lscape}
\usepackage{amsmath}
\usepackage{amssymb}
\usepackage{amsfonts}
\usepackage{graphicx}

\begin{document}

\title{Space-times over normed division algebras, revisited}
\author{R. Vilela Mendes\thanks{%
rvilela.mendes@gmail.com; rvmendes@fc.ul.pt; http://label2.ist.utl.pt/vilela/%
} \\
%EndAName
CMAFcIO and IPFN, Universidade de Lisboa}
\date{}
\maketitle

\begin{abstract}
Normed division and Clifford algebras have been extensively used in the past
as a mathematical framework to accommodate the structures of the standard
model and grand unified theories. Less discussed has been the question of
why such algebraic structures appear in Nature. One possibility could be an
intrinsic complex, quaternionic or octonionic nature of the spacetime
manifold. Then, an obvious question is why spacetime appears nevertheless to
be simply parametrized by the real numbers. How the real slices of an higher
dimensional spacetime manifold might be almost independent from each other
is discussed here. This comes about as a result of the different nature of
the representations of the real kinematical groups and those of the extended
spaces. Some of the internal symmetry transformations might however appear
as representations on homogeneous spaces of the extended group
transformations that cannot be implemented on the elementary states.
\end{abstract}

\section{Introduction}

The search for order has lead many authors to frame the standard model of
elementary particles as a representation of a transformation group on
Hilbert spaces based on higher division algebras, in particular octonionic
Hilbert spaces, although representations of exterior or Clifford algebras
might work as well \cite{Rowlands}. These ideas trace back to the works of G%
\"{u}naydin and Gursey (\cite{Gunaydin1}-\cite{Gunaydin4}) having been,
since then, further explored by many authors (\cite{Gursey}-\cite{Furey2}
and references therein). Higher division algebras also appear in
supersymmetry and superstring theories (\cite{Kugo}-\cite{Evans}) and
provide some clues on how to correct the shortcomings of grand unified
theories \cite{Furey1} \cite{Baez-GUT}.

A question, that also intrigued several authors in the past, is what could
be at the origin of the apparent relevance of these algebras to the
structure of the matter states. Could it be that spacetime itself has a
complex, quaternionic of even octonionic structure? Several authors (\cite%
{Newman1}-\cite{Brown}) have studied complex Minkowski spaces obtaining for
example new solutions of the field equations and hints on quantized
structures in space-time. Particularly interesting is the interpretation of
Minkowski space as a family of lines in $\mathbb{C}P^{3}$ \cite{Newman3}-%
\cite{Penrose3}. Nevertheless a question that remains open is why the
spacetime of ordinary life looks like a real space instead of one with
higher division algebra coordinates. Rather than the use of normed division
algebras as a framework to accomodate particle states and internal quantum
numbers, this is the main question that will be addressed in this paper

From a mathematical point of view, the idea of coordinates taken from a
higher division algebra might make sense. For the mathematical
reconstruction of space-time, it would be natural to start from the
empirical evidence for a four-dimensional structure with a pseudo-Euclidean
metric. Given this apriori fact, what else is needed to construct a
quantitative framework to describe the natural phenomena? At a minimum, we
must equip each one of the four dimensions with a numerical labelling. To
allow for the usual algebraic operations and the measurement of distances,
the labelling must, at least, be taken from a normed division algebra.
Hurwitz theorem now implies that the only labelling possibilities are the
reals, the complex, the quaternions and the octonions. Could it be that
spacetime is indeed a manifold better labelled by an higher division algebra
than by the real quantities of our everyday experience? If so, why do we not
feel it?

When, for the labeling of the dimensional coordinates, an algebra other than
the reals is used, consistency with the usual physical observations should
of course be mantained. Fixing in the extended space four independent
directions and using them as a basis for a real vector space one would
obtain \textit{a real slice of the extended space}. The consistency
condition is that the symmetry group of the extended space must reduce to
the real Lorentz and Poincar\'{e} groups in each real slice. Taking $%
\{e_{\nu }\}$ as an orthonormal set in one of the real slices, the set of
all real slices is spanned by 
\begin{equation*}
\left\{ e_{\mu }^{^{\prime }}\right\} =\left\{ \varphi (\mu )\Lambda _{\mu
}^{\nu }e_{\nu }\right\}
\end{equation*}%
where $\Lambda $ is a real Lorentz matrix and $\varphi (\mu )$ a $\mu -$%
dependent unit element of the chosen normed algebra $\mathbb{K}$, $\left(
\varphi (\mu )\in \mathbb{K},\left\vert \varphi (\mu )\right\vert =1\right) $%
. In each real slice $\left\vert x_{\mu }\right\vert g^{\mu \nu }\left\vert
x_{\nu }\right\vert $ is to be preserved and this implies that the Lorentz
invariance group in the extended space is 
\begin{equation}
\Lambda ^{\dagger }G\Lambda =G,  \label{I.1}
\end{equation}%
$G$ being the metric $(1,-1,-1,-1)$ and $\dagger $ the adjoint operation.
This would be the $U\left( 1,3,\mathbb{K}\right) $ group over the normed
algebra $\mathbb{K}$. Together with the inhomogeneous translations, these
are the groups that in the past, for $\mathbb{K=C}$, have been called the
Poincar\'{e} groups with real metric \cite{Barut} \cite{Ungar} \cite%
{Kursunoglu}. Notice that for $\mathbb{K=C}$ this is a $16$ parameter
complex Lorentz group distinct from the $12$ parameter group%
\begin{equation}
\Lambda ^{T}G\Lambda =G  \label{I.2}
\end{equation}%
used in the analytical continuation of the S-matrix and interpretation of
complex angular momentum \cite{Roff1} \cite{Roff2}.

In this paper the Poincar\'{e} groups with real metric will be studied over
the complex numbers, the quaternions and the octonions. Taking seriously the
hyphotesis that the true labelling of spacetime is an algebra larger than
the reals, one must of course understand why the spacetime of ordinary life
looks like a real space, that is, why the real slices look disjoint or
almost disjoint from one another. One possible (mathematical) answer is
somewhat surprising. It is based on the fact that not all representations of
the kinematical group of the real slices are present in the larger group and
when they exist there is a superselection rule operating between them. The
fact that half-integer elementary spin states cannot be "rotated" away from
a real slice has as a corollary a conversion of kinematical transformations
into internal symmetries on the fibers over an homogeneous space.

\section{Complex space-time}

\subsection{The complex Poincar\'{e} group}

The group of four-dimensional space-time with complex coordinates that, when
acting inside each real slice, reduces to the Lorentz group is $U(1,3,%
\mathbb{C})$ satisfying 
\begin{equation}
\Lambda ^{\dagger }G\Lambda =G  \label{2.1}
\end{equation}%
Adding the complex space-time translations one obtains the semi-direct
product 
\begin{equation}
T_{4}\circledS U(1,3,\mathbb{C})  \label{2.2}
\end{equation}%
a $24-$parameter group. The generators of its Lie algebra are $\left\{
M_{\mu \nu },N_{\mu \nu },K_{\mu },H_{\mu }\right\} $\footnote{$\mu ,\nu \in
\left\{ 0,1,2,3,4\right\} $%
\par
$i,j\in \left\{ 1,2,3\right\} $}:

\# $M_{\mu \nu }=-M_{\nu \mu }$ ($6$ generators) corresponding to the
transformations%
\begin{equation}
M_{ij}:\left\{ 
\begin{array}{lll}
x^{^{\prime }i} & = & x^{i}\cos \theta +x^{j}\sin \theta \\ 
x^{^{\prime }j} & = & -x^{i}\sin \theta +x^{j}\cos \theta%
\end{array}%
\right. ;\;M_{0i}:\left\{ 
\begin{array}{lll}
x^{^{\prime }0} & = & x^{0}\cosh u+x^{i}\sinh u \\ 
x^{^{\prime }i} & = & x^{0}\sinh u+x^{i}\cosh u%
\end{array}%
\right.  \label{2.3}
\end{equation}

\# $N_{\mu \nu }=N_{\nu \mu }$ ($10$ generators) corresponding to the
transformations\footnote{%
For $\mathbb{K=Q}$ or $\mathbb{O}$, $i$ should be replaced by imaginary
units in that algebra.}%
\begin{eqnarray}
\underset{i\neq j}{N_{ij}} &:&\left\{ 
\begin{array}{lll}
x^{^{\prime }i} & = & x^{i}\cos \theta +ix^{j}\sin \theta \\ 
x^{^{\prime }j} & = & ix^{i}\sin \theta +x^{j}\cos \theta%
\end{array}%
\right. ;\;N_{0i}:\left\{ 
\begin{array}{lll}
x^{^{\prime }0} & = & x^{0}\cosh u+ix^{i}\sinh u \\ 
x^{^{\prime }i} & = & -ix^{0}\sinh u+x^{i}\cosh u%
\end{array}%
\right.  \notag \\
N_{00} &:&\left\{ 
\begin{array}{lll}
x^{^{\prime }0} & = & e^{-i2\theta }x^{0}%
\end{array}%
\right. ;\;N_{ii}:\left\{ 
\begin{array}{lll}
x^{^{\prime }i} & = & e^{i2\theta }x^{i}%
\end{array}%
\right.  \label{2.4}
\end{eqnarray}

\# $K_{\mu }$ ($4$ generators) corresponding to the transformations%
\begin{equation}
K_{\mu }:\left\{ 
\begin{array}{lll}
x^{^{\prime \mu }} & = & x^{\mu }+\theta%
\end{array}%
\right.  \label{2.5}
\end{equation}

\# $H_{\mu }$ ($4$ generators) corresponding to the transformations%
\begin{equation}
H_{\mu }:\left\{ 
\begin{array}{lll}
x^{^{\prime \mu }} & = & x^{\mu }+i\theta%
\end{array}%
\right.  \label{2.6}
\end{equation}%
The commutation relations are%
\begin{eqnarray}
\left[ M_{\mu \nu },M_{\rho \sigma }\right] &=&-M_{\mu \sigma }g_{\nu \rho
}-M_{\nu \rho }g_{\mu \sigma }+M_{\nu \sigma }g_{\rho \mu }+M_{\mu \rho
}g_{\nu \sigma }  \notag \\
\left[ M_{\mu \nu },N_{\rho \sigma }\right] &=&-N_{\mu \sigma }g_{\nu \rho
}+N_{\nu \rho }g_{\mu \sigma }+N_{\nu \sigma }g_{\rho \mu }-N_{\mu \rho
}g_{\nu \sigma }  \notag \\
\left[ N_{\mu \nu },N_{\rho \sigma }\right] &=&M_{\mu \sigma }g_{\nu \rho
}+M_{\nu \rho }g_{\mu \sigma }+M_{\nu \sigma }g_{\rho \mu }+M_{\mu \rho
}g_{\nu \sigma }  \notag \\
\left[ M_{\mu \nu },K_{\rho }\right] &=&-g_{\nu \rho }K_{\mu }+g_{\mu \rho
}K_{\nu }  \notag \\
\left[ M_{\mu \nu },H_{\rho }\right] &=&-g_{\nu \rho }H_{\mu }+g_{\mu \rho
}H_{\nu }  \notag \\
\left[ N_{\mu \nu },K_{\rho }\right] &=&-g_{\nu \rho }H_{\mu }-g_{\mu \rho
}H_{\nu }  \notag \\
\left[ N_{\mu \nu },H_{\rho }\right] &=&g_{\nu \rho }K_{\mu }+g_{\mu \rho
}K_{\nu }  \notag \\
\left[ K_{\mu },H_{\rho }\right] &=&0  \label{2.7}
\end{eqnarray}

The structure of this group and some of its little groups were first studied
by Barut \cite{Barut}. Here one emphasizes its representations and, in
particular, the relation between the representations of the full group and
those of the real Poincar\'{e} group that operates on each real slice. The
subgroup generated by $\left\{ M_{\mu \nu },K_{\mu }\right\} $ is one of the
subgroups isomorphic to the real Poincar\'{e} group and $\sum_{\mu }N_{\mu
\mu }g_{\mu \mu }$ generates an invariant subgroup that commutes with all
the generators.

The invariants of the group are \cite{Barut}:%
\begin{eqnarray}
P^{2} &=&K^{\mu }K_{\mu }+H^{\mu }H_{\mu }  \notag \\
C_{3} &=&\left( K^{\mu }K_{\mu }+H^{\mu }H_{\mu }\right) N_{\mu }^{\mu
}-N_{\mu \nu }\left( K^{\mu }K^{\nu }+H^{\mu }H^{\nu }\right) -2M_{\mu \nu
}K^{\mu }H^{\nu }  \notag \\
C_{4} &=&\left( K^{\mu }K_{\mu }+H^{\mu }H_{\mu }\right) \left\{ \left(
M_{\mu \alpha }K^{\alpha }+N_{\mu \alpha }H^{\alpha }\right) \left( M^{\mu
\alpha }K_{\alpha }+N^{\mu \alpha }H_{\alpha }\right) \right.  \notag \\
&&\left. +\left( M_{\mu \alpha }H^{\alpha }-N_{\mu \alpha }K^{\alpha
}\right) \left( M^{\mu \alpha }H_{\alpha }-N^{\mu \alpha }K_{\alpha }\right)
\right\} -\frac{1}{2}\left\{ 2M_{\mu \nu }K^{\mu }H^{\nu }-N_{\mu \nu
}\left( K^{\mu }K^{\nu }+H^{\mu }H^{\nu }\right) \right\} ^{2}  \notag \\
&&-\frac{1}{2}\left( K^{\mu }K_{\mu }+H^{\mu }H_{\mu }\right) \left( M^{\mu
\nu }M_{\mu \nu }+N^{\mu \nu }N_{\mu \nu }\right)  \label{2.8}
\end{eqnarray}

This complex Lorentz group is simply connected. Therefore parity and time
reversal are continuously connected to the identity.

For convenience, in the construction of the little group representations,
the generators are further decomposed into%
\begin{equation}
M_{ij}=R_{k};\;M_{0i}=L_{i};\;N_{ij}=U_{k};\;N_{0i}=M_{i};\;N_{\mu \mu
}=-2g_{\mu \nu }C_{\mu }  \label{2.9}
\end{equation}%
with the usual permutation order being implied in the definitions of $R_{k}$
and $U_{k}$. A matrix representation of these generators for the complex,
the quaternionic and the octonionic cases is contained in Appendix A.

To study the irreducible representations of the inhomogeneous group (\ref%
{2.2}) the induced representation method is used. The translation generators
are diagonalized 
\begin{equation}
P_{\mu }\left\vert p\right\rangle =\left( K_{\mu }+H_{\mu }\right)
\left\vert p\right\rangle =\left( \textnormal{Re}p_{\mu }+i\textnormal{Im}p_{\mu
}\right) \left\vert p\right\rangle  \label{2.10}
\end{equation}%
and the little groups are classified according to the values $M_{0}^{2}$ of
the invariant $P^{2}=K^{2}+H^{2}$ \cite{Barut}.

(i) $M_{0}^{2}>0$. $p$ can be brought to the form $(p^{0},0,0,0)$ and the
little group $G_{1}^{\mathbb{C}}$ is a $U(3)$ subgroup generated by $\left\{
R_{i},U_{i},C_{i},i=1,2,3\right\} $

(ii) $M_{0}^{2}=0$ , $p^{\mu }\neq 0$. $p$ can be brought to the form $%
\left( p^{0},0,0,p^{0}\right) $. The little group $G_{2}^{\mathbb{C}}$ ,
generated by $\left\{
R_{3},U_{3},C_{1},C_{2},L_{1}+R_{2},L_{2}-R_{1},M_{1}+U_{2},M_{2}+U_{1},M_{3}+C_{3}-C_{0}\right\} 
$ is a semidirect product $N\circledS U(2)$ where $U(2)$ is generated by $%
\left\{ R_{3},U_{3},C_{1},C_{2}\right\} $ and the invariant subgroup $N$ by
the remaining generators.

(iii) $M_{0}^{2}=0$ , $p^{\mu }=0$. In this case the little group is the
full $U(1,3)$.

(iv) $M_{0}^{2}<0$ . $p$ can be brought to the form $\left(
0,0,0,p^{3}\right) $ and the little group is $U(1,2)$ generated by $\left\{
R_{3},L_{1},L_{2},U_{3},M_{1},M_{2},C_{1},C_{2},C_{0}\right\} $

Now the representations of the first two classes will be analysed.

\subsection{$M_{0}^{2}>0$ representations}

The little group that classifies the representations in this case is $%
U\left( 3\right) $ with hermitean generators%
\begin{equation}
-iR_{i};\;-iU_{i};\;-i\left( C_{1}-C_{2}\right) ;\;-i\left(
C_{1}+C_{2}-2C_{3}\right)  \label{C1}
\end{equation}%
generating $SU\left( 3\right) $ and a $U\left( 1\right) $ generator $%
-i\left( C_{1}+C_{2}+C_{3}\right) $. Notice that in this $u\left( 3\right) $
algebra the $\left\{ R_{i}\right\} $-subalgebra is the ordinary space
rotation algebra.

In the $4\times 4$ matrices representing $R_{i},U_{i}$ and $C_{i}$
(AppendixA), by suppressing the zero-valued first line and first column one
obtains the following correspondence to the usual $\lambda $ matrices of $%
su\left( 3\right) $%
\begin{eqnarray}
-iR_{1} &\rightarrow &\lambda _{7};\;-iR_{2}\rightarrow -\lambda
_{5};\;-iR_{3}\rightarrow \lambda _{2};\;\;-iU_{1}\rightarrow \lambda
_{6};\;\;-iU_{2}\rightarrow \lambda _{4};\;\;-iU_{3}\rightarrow \lambda _{1};
\notag \\
-i\left( C_{1}-C_{2}\right) &\rightarrow &\lambda _{3};\;-i\left(
C_{1}+C_{2}-2C_{3}\right) \rightarrow \sqrt{3}\lambda _{8};\;-i\left(
C_{1}+C_{2}+C_{3}\right) \rightarrow \boldsymbol{1}  \label{C2}
\end{eqnarray}%
All the irreducible representations of $SU\left( 3\right) $ are obtained by
tensor products of $\boldsymbol{3}$ and $\overline{\boldsymbol{3}}$. Hence
the spin operator acting on this states is%
\begin{equation}
J^{2}=-R_{1}^{2}-R_{2}^{2}-R_{3}^{2}=\lambda _{7}^{2}+\lambda
_{5}^{2}+\lambda _{2}^{2}=\left( 
\begin{array}{ccc}
2 & 0 & 0 \\ 
0 & 2 & 0 \\ 
0 & 0 & 2%
\end{array}%
\right) =\left( 1\times \left( 1+1\right) \right) \boldsymbol{1}_{3}
\label{C3}
\end{equation}%
which operating on $\boldsymbol{3}$ or $\overline{\boldsymbol{3}}$ yields
spin $1$. Therefore the conclusion is that only integer spins occur in the
reduction of $SU\left( 3\right) $ with respect to the rotation subgroup.

The extension of the kinematical symmetry group from the real to the complex
Poincar\'{e} group does not contain massive half-integer spin elementary
states. A phase generated by $-i\left( C_{1}+C_{2}+C_{3}\right) $, a $%
U\left( 1\right) $ quantum number, appears naturally. Kinematic group
transformations are the tools that, when applied to a state, exhibit all its
apects. In particular, general complex Lorentz transformations would take
states from one real slice to another. The fact that massive half-integer
spin states are not representations of the complex Poincar\'{e} group means
that these elementary states cannot be taken out of the real slice where
they are representations of the real Poincar\'{e} group. As an illustration
let us try to implement the complex Lorentz group $\left\{ M_{\mu \nu
},N_{\mu \nu }\right\} $ on a Dirac spinor.

The tensor structure of the Dirac spinor being coded in the Dirac equation,
the representation matrices $S$ of $M_{\mu \nu }$ and $N_{\mu\nu }$
must satisfy%
\begin{equation}
\Lambda _{\mu }^{\nu }\gamma ^{\mu }=S^{-1}\gamma ^{\nu }S  \label{C4}
\end{equation}%
with, for an infinitesimal transformation%
\begin{equation}
\Lambda _{\mu }^{\nu }=g_{\mu }^{\nu }+\Delta \omega _{\mu }^{\nu }
\label{C5}
\end{equation}%
and 
\begin{equation}
\Delta \omega ^{\nu \mu \ast }=-\Delta \omega ^{\mu \nu }  \label{C6}
\end{equation}%
following from (\ref{2.1}), which implies $\Delta \omega ^{\nu \mu }=-\Delta
\omega ^{\mu \nu }$ for $M_{\mu \nu }$ and $\Delta \omega ^{\nu \mu }=\Delta
\omega ^{\mu \nu }$ for $N_{\mu \nu }$.

Therefore, to implement the $6$ real Lorentz group transformations generated
by $R_{i}$ and $L_{i}$ ($M_{\mu \nu }$) in the Dirac equation one has to
find a functional $\Gamma _{\alpha \beta }$ of the gamma matrices satisfying%
\begin{equation*}
2i\left( g_{\alpha }^{\nu }\gamma _{\beta }-g_{\beta }^{\nu }\gamma _{\alpha
}\right) =\left[ \gamma ^{\nu },\Gamma _{\alpha \beta }\right]
\end{equation*}%
which has the solution%
\begin{equation*}
\Gamma _{\alpha \beta }=\frac{i}{2}\left[ \gamma _{\alpha },\gamma _{\beta }%
\right]
\end{equation*}%
To implement the remaning $10$ generators ($N_{\mu \nu }$) the corresponding
equation would be%
\begin{equation*}
2i\left( g_{\alpha }^{\nu }\gamma _{\beta }+g_{\beta }^{\nu }\gamma _{\alpha
}\right) =\left[ \gamma ^{\nu },N_{\alpha \beta }\right]
\end{equation*}%
with $N_{\alpha \beta }$ a symmetric functional of the gamma matrices. But
because the only such symmetric functional is $g_{\alpha \beta }$, the
equation has no solution.

The fact that half-integer spin states cannot be elementary states of the
complex Poincar\'{e} group, only elementary states of the real group,
implies that matter composed of half-integer elementary blocks cannot
communicate between different real slices, in the sense that they cannot be
rotated from one real slice to another. Half-integer spins might still be
associated to the SU(3) group in a nonlinear sense through an induced
representation on a homogeneous space. However rather than rotating the
states away from the real slice a multiplicity of identical representation
spaces is obtained. This is discussed in Appendix D.

By contrast with half-integer states, integer spin states may be bona-fide
elementary states of the complex group. However, these states are of a
special nature. In the complex Lorentz group parity and time-reversal are
continuously connected to the identity. Therefore in faithfull continuous
norm-conserving representations of this group both parity and time reversal
must be implemented by unitary operators. On the other hand because of
energy positivity \cite{Halpern}, the time reversal operation in a state $%
\psi _{R}\in V_{R}$ of the real Poincar\'{e} group must be implemented by an
antiunitary operator. Therefore between these "real slice states" $\psi _{R}$
and those that are faithfull representations of the complex group $\psi
_{C}\in V_{C}$, there is a superselection rule. Consider a linear
superposition of two of these states%
\begin{equation*}
\Phi =\alpha \psi _{R}+\beta \psi _{C}
\end{equation*}%
with $\alpha ,\beta $ reals numbers. Now $\Phi $ and $e^{i\theta }\Phi
=\alpha e^{i\theta }\psi _{R}+\beta e^{i\theta }\psi _{C}$ belong to the
same ray and therefore should represent the same state. Applying the time
reversal operator to both $\Phi $ and $e^{i\theta }\Phi $%
\begin{eqnarray*}
T\Phi &=&\alpha T\psi _{R}+\beta T\psi _{C} \\
Te^{i\theta }\Phi &=&\alpha e^{-i\theta }T\psi _{R}+\beta e^{i\theta }T\psi
_{C}
\end{eqnarray*}%
$T\Phi $ and $Te^{i\theta }\Phi $ belong to different rays, hence $T$ does
not establish a ray correspondence in $V_{R}\oplus V_{C}$ unless $\alpha =0$
or $\beta =0$, that is, $V_{R}$ and $V_{C}$ belong to different
superselection sectors. The fact that the integer spin states in $V_{C}$ are
in a different superselection sector does not mean that they cannot interact
with the states in $V_{R}$. Whereas the superselection rule result concerns
the structure of the direct sum $V_{R}\oplus V_{C}$, the nature of the
interactions depends on the way the group transformations operate in the
tensor product $V_{R}\otimes V_{C}$. Consider now the $T$ operation acting
on $V_{R}\otimes V_{C}$ and compute its action on a matrix element%
\begin{eqnarray*}
\left( T\left( \psi _{R}^{(1)}\otimes \psi _{C}^{(1)}\right) ,T\left( \psi
_{R}^{(2)}\otimes \psi _{C}^{(2)}\right) \right) &=&\left( \psi
_{R}^{(2)}\otimes \psi _{R}^{(1)}\right) \left( \psi _{C}^{(1)}\otimes \psi
_{C}^{(2)}\right) \\
&=&\left( \psi _{R}^{(2)}\otimes \psi _{C}^{(1)},\psi _{R}^{(1)}\otimes \psi
_{C}^{(2)}\right)
\end{eqnarray*}%
Therefore there is no choice of phases that can make $T$ a unitary or
anti-unitary operator in the tensor product space. Therefore by Wigner's
theorem $T$ cannot be a symmetry in $V_{R}\otimes V_{C}$ \footnote{%
A similar situation has been discussed in the past \cite{VilelaJPA} for
states transforming under a different group, the analytical continuation
complex Lorentz group $\left( \Lambda ^{T}G\Lambda =G\right) $, which also
connects the identity to time reversal although not to parity.}.

In conclusion: Half-integer spin states and integer spin states with
antiunitary time reversal transformations are confined to one real slice.
Integer spin states of the full complex Poincar\'{e} group, when interacting
with the inner states of the real slices can only mediate T-violating
interactions (CP violation or possibly gravitational interactions \cite%
{Barbour} \cite{Carlip} \cite{Albrecht} \cite{Carroll}).

\subsection{$M_{0}^{2}=0,$ $p^{\protect\mu }\neq 0$ representations}

For the $M_{0}^{2}=0,$ $p^{\mu }\neq 0$ case the algebra of the little group 
$G_{2}^{\mathbb{C}}$ for the momentum $\left( p^{0},0,0,p^{0}\right) $ is a
semi-direct sum%
\begin{equation*}
\mathcal{L}G_{2}^{\mathbb{C}}=N^{c}\Diamond H^{c}
\end{equation*}%
($\left[ N^{c},N^{c}\right] \subset N^{c};\;\left[ H^{c},N^{c}\right]
\subset N^{c};\;\left[ H^{c},H^{c}\right] \subset H^{c}$), $N^{c}$, the
algebra of the normal subgroup $\mathcal{N}^{c}$, being a 2-dimensional
Heisenberg algebra%
\begin{equation*}
N^{c}=\left\{
l_{1}=L_{1}+R_{2};m_{1}=M_{1}+U_{2};l_{2}=L_{2}-R_{1};m_{2}=M_{2}+U_{1};m_{3}=M_{3}+C_{3}-C_{0}\right\}
\end{equation*}%
and $H^{c}$ is the algebra of $\mathcal{H}^{c}=U\left( 2\right) $%
\begin{equation*}
H^{c}=\left\{ R_{3};U_{3};\frac{1}{2}\left( C_{1}-C_{2}\right)
;C_{1}+C_{2}\right\}
\end{equation*}%
The representations of $G_{2}^{\mathbb{C}}$ are also obtained by the induced
representation method. Given a representation $\alpha $ of $\mathcal{N}^{c}$
one finds the isotropy subgroup $\mathcal{H}^{c}\left( \alpha \right) $ of $%
\mathcal{H}^{c}$, that is,%
\begin{equation*}
\alpha \left( hnh^{-1}\right) =\alpha \left( n\right) \hspace{1cm}\forall
n\in \mathcal{N}^{c},\forall h\in \mathcal{H}^{c}
\end{equation*}%
Then, for each representation $\beta $ of $\mathcal{H}^{c}\left( \alpha
\right) $, the product representation $\sigma =\alpha \times \beta $ defines
a homogeneous vector bundle over $G_{2}^{c}/\mathcal{N}^{c}\mathcal{H}^{c}$. 
$\mathcal{N}^{c}\mathcal{H}^{c}$ acts on the fiber by the $\sigma $
representation and the full $G_{2}^{c}$ representation is obtained by
composing $\sigma $ with the translations in the base manifold $G_{2}^{c}/%
\mathcal{N}^{c}\mathcal{H}^{c}$. The states in the representation of $%
G_{2}^{c}$ are labelled by a point in the base manifold $G_{2}^{c}/\mathcal{N%
}^{c}\mathcal{H}^{c}$ and by the fiber indices of the $\sigma $
representation. To classify the representations of $G_{2}^{c}$ all one has
to do is to classify the $\sigma -$representations of $\mathcal{N}^{c}%
\mathcal{H}^{c}$ for each $\alpha $.

The generator $m_{3}$ commutes with all generators in $\mathcal{L}G_{2}^{c}$%
. Therefore it is a constant in each irreducible representation. By the
Stone-von Neumann theorem this constant uniquely characterizes an
irreducible representation of $\mathcal{N}^{c}$, namely, for the algebra
elements%
\begin{eqnarray*}
l_{1}\psi ^{\left( \mu \right) }\left( \eta ,\xi \right) &=&i\eta \psi
^{\left( \mu \right) }\left( \eta ,\xi \right) \\
l_{2}\psi ^{\left( \mu \right) }\left( \eta ,\xi \right) &=&i\xi \psi
^{\left( \mu \right) }\left( \eta ,\xi \right) \\
m_{1}\psi ^{\left( \mu \right) }\left( \eta ,\xi \right) &=&-\mu \frac{%
\partial }{\partial \eta }\psi ^{\left( \mu \right) }\left( \eta ,\xi \right)
\\
m_{2}\psi ^{\left( \mu \right) }\left( \eta ,\xi \right) &=&-\mu \frac{%
\partial }{\partial \xi }\psi ^{\left( \mu \right) }\left( \eta ,\xi \right)
\\
m_{3}\psi ^{\left( \mu \right) }\left( \eta ,\xi \right) &=&\frac{i}{2}\mu
\psi ^{\left( \mu \right) }\left( \eta ,\xi \right)
\end{eqnarray*}%
realized in the space of functions of two real variables $\eta $ and $\xi $.
Therefore there are two types of representations of $G_{2}^{c}$.

$\left( 1\right) $ For a nontrivial $\alpha -$representation of $\mathcal{N}%
^{c} $ of the type above, the little group $\mathcal{H}^{c}\left( \alpha
\right) $ is empty. Therefore $\beta $ is trivial and the representations
are simply labelled by the differentiable functions $\psi ^{\left( \mu
\right) }\left( \eta ,\xi \right) $. This is analogous to the continuous
spin group of the real Poincar\'{e} group.

$\left( 2\right) $ If $\alpha $ is trivial, that is, if the generators $%
l_{i},m_{i}$ are mapped on the zero operator, then $\mathcal{H}^{c}\left(
\alpha \right) =U\left( 2\right) $. The states are now labelled by the
quantum numbers of $SU\left( 2\right) $ and a $U\left( 1\right) $ phase.
Notice however that each spin projection in a $SU\left( 2\right) $ multiplet
may correspond to a different particle in the "real slice" because only $%
R_{3}$ among the $SU\left( 2\right) $ generators belongs to the real Lorentz
group.

As seen from the commutation table in Appendix C the normalized generators
of the $SU\left( 2\right) $ subgroup, that label the states, are $\left\{ 
\frac{R_{3}}{2},\frac{U_{3}}{2},\frac{1}{2}\left( C_{1}-C_{2}\right)
\right\} $. Therefore because in the representations of $SU\left( 2\right) $%
, $\frac{R_{3}}{2}$ has an integer or half-integer spectrum, the spectrum of 
$R_{3}$ is integer and, once again, one finds that half-integer spins states
cannot have the full complex Poincar\'{e} group as a symmetry group.

The fact that for the massless case half-integer spin states cannot be
elementary states of the complex Poincar\'{e} group, only elementary states
of the real group, implies, as in the massive case, that matter composed of
half-integer elementary blocks cannot communicate between real slices.
Integer spin states may be bona-fide elementary states of the complex group.
However, as seen before, their interactions with other states in each real
slice still depend on the way the discrete transformations are implemented,
leading, as discussed before to a superselection rule and T-violation.

The lowest massless spin multiplets that naturally appear on the complex
Poincar\'{e} group are a zero spin singlet, a spin $1$ with $+1$ and $-1$
projections and a multiplet containing $+2$ and $-2$ as well as a scalar.

\section{Quaternionic and octonionic space-times}

The generators of the quaternionic and octonionic Lorentz algebras ($\mathbb{%
K=Q}$ or $\mathbb{O}$) are%
\begin{equation*}
\left\{ R_{a};L_{a};U_{a}^{(\alpha )};M_{a}^{(\alpha )};C_{a}^{\left( \alpha
\right) };C_{0}^{\left( \alpha \right) }\right\}
\end{equation*}%
with $a=1,2,3$ and $\alpha =i,j,k$, for the imaginary quaternionic units, $%
i^{2}=j^{2}=k^{2}=-1;ij=-ji;ij=k$ and cyclic permutations and $\alpha =e_{1}$
to $e_{7}$ for the octonions. A matrix representation of these generators is
contained in the Appendix A.

The quaternionic Lorentz group has $36$ generators and, with the $4$
quaternionic translations, the corresponding Poincar\'{e} group has $40$
generators. The octonionic Lorentz algebra has $76$ generators which
together with $8$ translations adds up to $84$ generators. It is an algebra
closed under commutation, however not a Lie algebra because of the
non-associativity of the octonions.

As for the complex Poincar\'{e} group one analyses the two cases: $M^{2}>0$
and $M^{2}=0$, $p^{\mu }\neq 0$.

\subsection{$M_{0}^{2}>0$ representations}

Bringing \ $p$ to the form $\left( p^{0},0,0,0\right) $ the little algebra, $%
G_{2}$, is generated by $21$ generators for the quaternionic case and $45$
generators for the octonionic case.%
\begin{equation*}
\left\{ R_{i};U_{i}^{\left( e\right) };C_{1}^{\left( e\right)
};C_{2}^{\left( e\right) };C_{3}^{\left( e\right) }\right\}
\end{equation*}%
with the commutation table of Appendix B. These algebras are $u\left( 3,%
\mathbb{Q}\right) $ and $u\left( 3,\mathbb{O}\right) $. Notice however that $%
u\left( 3,\mathbb{O}\right) $ is closed under commutation but not a Lie
algebra. For example:%
\begin{eqnarray}
&&\left[ U_{1}^{\left( e_{1}\right) },\left[ U_{2}^{\left( e_{2}\right)
},U_{3}^{\left( e_{6}\right) }\right] \right] +\left[ U_{2}^{\left(
e_{2}\right) },\left[ U_{3}^{\left( e_{6}\right) },U_{1}^{\left(
e_{1}\right) }\right] \right] +\left[ U_{3}^{\left( e_{6}\right) },\left[
U_{1}^{\left( e_{1}\right) },U_{2}^{\left( e_{2}\right) }\right] \right] 
\notag \\
&=&-\left[ U_{1}^{\left( e_{1}\right) },U_{1}^{\left( e_{4}\right) }\right] +%
\left[ U_{2}^{\left( e_{2}\right) },U_{2}^{\left( e_{7}\right) }\right] +%
\left[ U_{3}^{\left( e_{6}\right) },U_{3}^{\left( e_{3}\right) }\right]
=-4\left( C_{1}+C_{2}+C_{3}\right) ^{\left( e_{5}\right) }  \label{Q1}
\end{eqnarray}%
For $M_{0}^{2}>0$ one has in all cases (including the complex) the algebras $%
u\left( 3,\mathbb{K}\right) $ with $\mathbb{K=C},\mathbb{Q},\mathbb{O}$
generated by the skew-hermiteam matrices%
\begin{equation*}
M=\left\{ R_{i};U_{i}^{\left( e\right) };C_{1}^{\left( e\right)
};C_{2}^{\left( e\right) };C_{3}^{\left( e\right) }\right\}
\end{equation*}%
$M=-M^{\dag }$, where here only the $3\times 3$ block of the generators in
Appendix A are considered%
\begin{equation*}
R_{1}=\left( 
\begin{array}{lll}
0 & 0 & 0 \\ 
0 & 0 & 1 \\ 
0 & -1 & 0%
\end{array}%
\right) \;R_{2}=\left( 
\begin{array}{lll}
0 & 0 & -1 \\ 
0 & 0 & 0 \\ 
1 & 0 & 0%
\end{array}%
\right) \;R_{3}=\left( 
\begin{array}{lll}
0 & 1 & 0 \\ 
-1 & 0 & 0 \\ 
0 & 0 & 0%
\end{array}%
\right)
\end{equation*}%
\begin{equation*}
U_{1}^{\left( e\right) }=\left( 
\begin{array}{lll}
0 & 0 & 0 \\ 
0 & 0 & e \\ 
0 & e & 0%
\end{array}%
\right) \;U_{2}^{\left( e\right) }=\left( 
\begin{array}{lll}
0 & 0 & e \\ 
0 & 0 & 0 \\ 
e & 0 & 0%
\end{array}%
\right) \;U_{3}^{\left( e\right) }=\left( 
\begin{array}{lll}
0 & e & 0 \\ 
e & 0 & 0 \\ 
0 & 0 & 0%
\end{array}%
\right)
\end{equation*}%
\begin{equation}
C_{1}^{\left( e\right) }=\left( 
\begin{array}{lll}
e & 0 & 0 \\ 
0 & 0 & 0 \\ 
0 & 0 & 0%
\end{array}%
\right) \;C_{2}^{\left( e\right) }=\left( 
\begin{array}{lll}
0 & 0 & 0 \\ 
0 & e & 0 \\ 
0 & 0 & 0%
\end{array}%
\right) \;C_{3}^{\left( e\right) }=\left( 
\begin{array}{lll}
0 & 0 & 0 \\ 
0 & 0 & 0 \\ 
0 & 0 & e%
\end{array}%
\right)  \label{Q2}
\end{equation}

with $e\in \mathbb{K}^{\prime }$ (an imaginary element of $\mathbb{K}$). For
all these matrices $M=-M^{\dag }$, with $e^{\dag }=-e$, $e\in \mathbb{K}%
^{\prime }$.

In these algebras there is a $u\left( 1,\mathbb{K}\right) $ subalgebra
generated by $C_{1}^{\left( e\right) }+C_{2}^{\left( e\right)
}+C_{3}^{\left( e\right) }$, whereas $\left\{ R_{i};U_{i}^{\left( e\right)
};C_{1}^{\left( e\right) }-C_{2}^{\left( e\right) };C_{1}^{\left( e\right)
}+C_{2}^{\left( e\right) }-2C_{3}^{\left( e\right) }\right\} $ generates $%
su\left( 3,\mathbb{K}\right) $.

The root space decomposition may easily be generalized for these algebras.
Let%
\begin{eqnarray}
h_{1} &=&-e\left( C_{1}^{\left( e\right) }-C_{2}^{\left( e\right) }\right) 
\notag \\
h_{2} &=&\frac{-e}{\sqrt{3}}\left( C_{1}^{\left( e\right) }+C_{2}^{\left(
e\right) }-2C_{3}^{\left( e\right) }\right)  \label{Q3}
\end{eqnarray}%
Then%
\begin{equation}
\begin{array}{lll}
\left[ h_{1},\left( -eU_{3}^{\left( e\right) }+R_{3}\right) \right] =2\left(
-eU_{3}^{\left( e\right) }+R_{3}\right) & ; & \left[ h_{2},\left(
-eU_{3}^{\left( e\right) }+R_{3}\right) \right] =0 \\ 
\left[ h_{1},\left( -eU_{3}^{\left( e\right) }-R_{3}\right) \right]
=-2\left( -eU_{3}^{\left( e\right) }-R_{3}\right) & ; & \left[ h_{2},\left(
-eU_{3}^{\left( e\right) }-R_{3}\right) \right] =0 \\ 
\left[ h_{1},\left( -eU_{1}^{\left( e\right) }+R_{1}\right) \right] =-\left(
-eU_{1}^{\left( e\right) }+R_{1}\right) & ; & \left[ h_{2},\left(
-eU_{1}^{\left( e\right) }+R_{1}\right) \right] =\sqrt{3}\left(
-eU_{1}^{\left( e\right) }+R_{1}\right) \\ 
\left[ h_{1},\left( -eU_{1}^{\left( e\right) }-R_{1}\right) \right] =\left(
-eU_{1}^{\left( e\right) }-R_{1}\right) & ; & \left[ h_{2},\left(
-eU_{1}^{\left( e\right) }-R_{1}\right) \right] =-\sqrt{3}\left(
-eU_{1}^{\left( e\right) }-R_{1}\right) \\ 
\left[ h_{1},\left( -eU_{2}^{\left( e\right) }-R_{2}\right) \right] =\left(
-eU_{2}^{\left( e\right) }-R_{2}\right) & ; & \left[ h_{2},\left(
-eU_{2}^{\left( e\right) }-R_{2}\right) \right] =\sqrt{3}\left(
-eU_{2}^{\left( e\right) }-R_{2}\right) \\ 
\left[ h_{1},\left( -eU_{2}^{\left( e\right) }+R_{2}\right) \right] =-\left(
-eU_{2}^{\left( e\right) }+R_{2}\right) & ; & \left[ h_{2},\left(
-eU_{2}^{\left( e\right) }+R_{2}\right) \right] =-\sqrt{3}\left(
-eU_{2}^{\left( e\right) }+R_{2}\right)%
\end{array}
\label{Q4}
\end{equation}%
The roots are the same as in $su\left( 3,\mathbb{C}\right) $, the difference
being that the nonzero root spaces now have dimension equal to $\dim \mathbb{%
K}^{\prime }$, three for quaternions and seven for octonions. The root space
of the zero root has dimension $2\times \dim \mathbb{K}^{\prime }$.

Representations of $su\left( 3,\mathbb{Q}\right) $ and $su\left( 3,\mathbb{O}%
\right) $ in quaternionic and octonionic Hilbert spaces are simply obtained
from those of $su\left( 3,\mathbb{C}\right) $. Notice that $\left(
-eU_{3}^{\left( e\right) }\pm R_{3}\right) ,\left( -eU_{1}^{\left( e\right)
}\pm R_{1}\right) $ and $\left( -eU_{2}^{\left( e\right) }\pm R_{2}\right) $
are the same for all $e$. Therefore given a complex representation of $%
su\left( 3,\mathbb{C}\right) $ with vectors $\left\{ \psi _{a}^{\left(
\lambda \right) }\right\} $ for the highest weight $\lambda $%
\begin{equation}
\left( -eU_{i}^{\left( e\right) }\right) \psi _{a}^{\left( \lambda \right)
}=\sum_{b=1}^{\dim \lambda }c_{ba}\psi _{b}^{\left( \lambda \right) }
\label{Q5}
\end{equation}%
with $c_{ba}\in \mathbb{C}$, one obtains%
\begin{equation}
U_{i}^{\left( e\right) }\psi _{a}^{\left( \lambda \right) }=\sum_{b=1}^{\dim
\lambda }ec_{ba}\psi _{b}^{\left( \lambda \right) }  \label{Q6}
\end{equation}%
with matrix elements $ec_{ba}\in \mathbb{Q},\mathbb{O}$. The same procedure
applies for the other generators and this makes sense in the framework of
quaternionic or octonionic Hilbert spaces.

A different point of view would be to look for representations of these
algebras with complex matrices. This may be achieved by defining for each
state of the $\lambda -$representation $\dim \mathbb{K}^{\prime }-1$ new
states%
\begin{equation*}
\psi _{a,e}^{\left( \lambda \right) }\circeq e\psi _{a}^{\left( \lambda
\right) }
\end{equation*}%
and Eq.(\ref{Q6}) becomes%
\begin{equation*}
U_{i}^{\left( e\right) }\psi _{a}^{\left( \lambda \right) }=\sum_{b=1}^{\dim
\lambda }c_{ba}\psi _{b,e}^{\left( \lambda \right) }
\end{equation*}%
a representation with complex matrices in a space of dimension $\dim \lambda
\times \left( \dim \mathbb{K}^{\prime }-1\right) $. Notice that one is
idenfying $e_{1}$ with the complex imaginary unit. The complex
representations so obtained are not necessarily irreducible. The lowest
nontrivial representations would be six dimensional for quaternions and
eighteen dimensional for octonions, corresponding to two and six spin one
states. Although octonions are a non-associative algebra this procedure
allows for a consistent construction of a representation space for $su\left(
3,\mathbb{O}\right) $. However because of the non-associativity of octonions
the corresponding matrices are a representation of $su\left( 3,\mathbb{O}%
\right) $ only in the quasi-algebra sense \cite{Albu1} \cite{Albu2} \cite%
{Albu3} \cite{Panaite}.

In the quaternionic case $su\left( 3,\mathbb{Q}\right) $ is a Lie algebra
and its correspondence to an ordinary complex algebra may be explicitly
exhibited. Representing the imaginary quaternion units by sigma matrices $%
\left( e_{i}\rightarrow -i\sigma _{i}\right) $ the generators of $su\left( 3,%
\mathbb{Q}\right) $ become%
\begin{equation*}
R_{ii}=R_{i}\otimes \boldsymbol{1}_{2};\;U_{ij}=U_{i}\otimes \left( -i\sigma
_{j}\right) ;\;C_{ij}=C_{i}\otimes \left( -i\sigma _{j}\right)
\end{equation*}%
The matrices of the algebra $\mathcal{A}=\left\{
R_{ii},U_{ij},C_{ij}\right\} $ satisfy%
\begin{equation*}
A^{T}\Omega +\Omega A=0\;\;A\in \mathcal{A}
\end{equation*}%
where $\Omega $ is the symplectic form%
\begin{equation*}
\Omega =\boldsymbol{1}_{3}\otimes \left( i\sigma _{2}\right)
\end{equation*}%
Therefore the algebra is the algebra of the symplectic group in $6$
dimensions. With%
\begin{equation*}
\Omega _{12}=-\Omega _{21}=\Omega _{34}=-\Omega _{43}=\Omega _{56}=-\Omega
_{65}=1
\end{equation*}%
all other elements $\Omega _{ij}$ being zero, the irreducible
representations are obtained from (symplectic) traceless tensors%
\begin{equation*}
\Omega _{ij}F_{\cdots ij\cdots }=0
\end{equation*}%
of definite permutation symmetry corresponding to Young diagrams $\left(
\sigma _{1},\sigma _{2},\sigma _{3}\right) $ with at most three lines.

The lowest dimensional representations have dimensions $0$ $(000),6$ $\left(
100\right) ,14$ $\left( 110\right) $ and $\left( 111\right) ,21$ $\left(
200\right) ,\cdots $ . Homogeneous polynomial basis for the defining $\left(
6\right) $ and the adjoint $\left( 21\right) $ representations are $\left(
x_{i};i=1\cdots 6\right) $ and $\left( x_{i}^{2},x_{i}x_{j}\;i<j;i,j=1\cdots
6\right) $. As in the complex Poincar\'{e} case only integer spins exist in
the irreducible representations. In the $6-$representation there are two
independent spin one states associated to $\left( x_{1},x_{3},x_{5}\right) $
and $\left( x_{2},x_{4},x_{6}\right) $.

\subsection{$M_{0}^{2}=0,$ $p^{\protect\mu }\neq 0$ representations}

Here for definitness the quaternionic case will be analysed. As before,
bringing $p$ to the form $\left( p^{0},0,0,p^{0}\right) $, the little group $%
G_{2}^{q}$ is generated by the following $21$ generators%
\begin{equation*}
\left\{ R_{3};U_{3}^{\alpha };C_{1}^{\alpha };C_{2}^{\alpha
};L_{1}+R_{2};L_{2}-R_{1};M_{1}^{\alpha }+U_{2}^{\alpha };M_{2}^{\alpha
}+U_{1}^{\alpha };M_{3}^{\alpha }+C_{3}^{\alpha }-C_{0}^{\alpha }\right\}
\end{equation*}%
the commutation table for these generators being listed in the Appendix C.

From the commutation table one sees that the algebra of the little group $%
G_{2}^{q}$ is the semidirect sum%
\begin{equation*}
\mathcal{L}G_{2}^{q}=N^{q}\Diamond H^{q}
\end{equation*}%
$N^{q}=\left\{ l_{1},l_{2},m_{1}^{\alpha },m_{2}^{\alpha },m_{3}^{\alpha
}\right\} $ and $H^{q}=\left\{ R_{3},U_{3}^{\alpha },C_{1}^{\alpha
}-C_{2}^{\alpha },C_{1}^{\alpha }+C_{2}^{\alpha }\right\} $.

The generators $m_{3}^{\alpha }$ commute with all the other generators,
constants in an irreducible representation, are denoted $\frac{i}{2}\mu _{i},%
\frac{i}{2}\mu _{j},\frac{i}{2}\mu _{k}$. Then, from the commutation table
in Appendix C one concludes that the invariant algebra $N^{q}$ consists of a
set of overlapping Heisenberg algebras which, on the space of differentiable
functions of 6 variables $\overrightarrow{\eta }=\eta _{i},\eta _{j},\eta
_{k},\xi _{i},\xi _{j},\xi _{k}$ have the representation%
\begin{eqnarray*}
l_{1}\psi \left( \overrightarrow{\eta },\overrightarrow{\xi }\right)
&=&i\left( \eta _{i}+\eta _{j}+\eta _{k}\right) \psi \left( \overrightarrow{%
\eta },\overrightarrow{\xi }\right) \\
m_{1}^{i}\psi \left( \overrightarrow{\eta },\overrightarrow{\xi }\right)
&=&i\left( -\mu _{i}\frac{\partial }{\partial \eta _{i}}+i\frac{\mu _{k}}{%
\mu _{j}}\eta _{j}-i\frac{\mu _{j}}{\mu _{k}}\eta _{k}\right) \psi \left( 
\overrightarrow{\eta },\overrightarrow{\xi }\right) \\
m_{1}^{j}\psi \left( \overrightarrow{\eta },\overrightarrow{\xi }\right)
&=&i\left( -\mu _{j}\frac{\partial }{\partial \eta _{j}}+i\frac{\mu _{i}}{%
\mu _{k}}\eta _{k}-i\frac{\mu _{k}}{\mu _{i}}\eta _{i}\right) \psi \left( 
\overrightarrow{\eta },\overrightarrow{\xi }\right) \\
m_{1}^{k}\psi \left( \overrightarrow{\eta },\overrightarrow{\xi }\right)
&=&i\left( -\mu _{k}\frac{\partial }{\partial \eta _{k}}+i\frac{\mu _{j}}{%
\mu _{i}}\eta _{i}-i\frac{\mu _{i}}{\mu j}\eta _{j}\right) \psi \left( 
\overrightarrow{\eta },\overrightarrow{\xi }\right) \\
l_{2},m_{2}^{\alpha } &\rightarrow &\left( \eta \rightarrow \xi \right) \\
m_{3}^{\alpha }\psi \left( \overrightarrow{\eta },\overrightarrow{\xi }%
\right) &=&\frac{i}{2}\mu _{\alpha }\psi \left( \overrightarrow{\eta },%
\overrightarrow{\xi }\right)
\end{eqnarray*}%
As in the complex case there are two classes of representations

1 - For a nontrivial representation of $N^{q}$, as above, the little group $%
\mathcal{H}^{q}\left( \alpha \right) $ is empty and the states are labelled
by the functions $\psi \left( \overrightarrow{\eta },\overrightarrow{\xi }%
\right) $. It is the continuous spin case.

2 - For a trivial (identically zero) representation of $N^{q}$ the little
group is $\mathcal{H}^{q}$ itself. $H^{q}$ is the algebra of $Sp\left(
2\right) \sim SO\left( 5\right) $. A Cartan subalgebra is $\left\{
iR_{3},iC_{+}^{i}=i\left( C_{1}^{i}+C_{2}^{i}\right) \right\} $ and the root
vectors are 
\begin{eqnarray*}
&&iU_{3}^{j}-\varepsilon _{2}U_{3}^{k}-\varepsilon _{1}\left(
C_{-}^{j}+\varepsilon _{2}iC_{-}^{k}\right) \\
&&iC_{+}^{j}-\varepsilon C_{+}^{k} \\
&&iU_{3}^{i}-\varepsilon C_{-}^{i}
\end{eqnarray*}%
$\varepsilon _{1}$ and $\varepsilon _{2}$ are independent $\pm $ signs. The
first $4$ root vectors have weigths $2\left( 
\begin{array}{c}
\varepsilon _{1} \\ 
\varepsilon _{2}%
\end{array}%
\right) $ and the other $4$ have weigths $2\left( 
\begin{array}{c}
0 \\ 
\varepsilon%
\end{array}%
\right) $ and $2\left( 
\begin{array}{c}
\varepsilon \\ 
0%
\end{array}%
\right) $.

In conclusion: Both in the quternionic and octonionic cases, the same
restrictions as in the complex case, are obtained concerning the separation
of the states of the full group from those of the real Poincar\'{e} group in
the real slices. These are, as before, the nonexistence of half-integer
spins as elementary states and the superselection rule for interger spins.
The main difference from the complex case is the multiplicity of states,
which, of course, has implications on the nature of the hypothetical
conversion of kinematical into internal symmetries (see Appendix D).

\section{Conclusions}

A study of matter representations in spacetimes over the complex,
quaternions and octonions has been performed. The main conclusions are:

1) No elementary half-integer spin states exist consistent with the full
operations of this higher dimensional division algebras. Hence, no
"rotation" between real slices for half-integer elementary states. It would
explain why half-integer spin matter stays confined to a real slice,
essentially unware of the larger spacetime where it migh be embeded.

2) Integer spin states exist as elementary states of the larger groups.
However there is a superselection rule operating between these states and
those of the real slices, potentially leading to T-violating interactions.

3) When, in the nonlinear sense of Appendix D, half-integer spin states are
associated to the higher groups, rather than a transformation between real
slices, a multiplicity of states is generated. In this sense the
transformations of the higher dimensional kinematical group play the role of
internal symmetries.

4) These general conclusions apply to complex, quaternionic and octonionic
spaces, the differences being mostly on the multiplicity of states.

5) If indeed the actual structure of space-time is associated to the higher
dimensional division algebras with a foliation in real slices, an obvious
question would be how many of these real slices are populated by ordinary
matter? Is there a quantum like restriction on the density of real slices?
Is it related to quantization of non-commutative coordinates?

\section*{\textbf{APPENDIX A}}

\textbf{A matrix representation of the generators of the complex,
quaternionic and octonionic Lorentz transformations}

From 
\begin{equation}
\left( 1+\omega ^{\dagger }\right) G\left( 1+\omega \right) =G  \label{A.1}
\end{equation}%
it follows 
\begin{equation}
\omega _{\sigma \mu }^{\ast }g_{\sigma \nu }=-g_{\mu \sigma }\omega _{\sigma
\nu }  \label{A.2}
\end{equation}%
and a set of independent generators is obtained from the conditions $\omega
_{00}^{\ast }=-\omega _{00}$ , $\omega _{k0}^{\ast }=\omega _{0k}$ and $%
\omega _{ik}^{\ast }=-\omega _{ki}$ 
\begin{equation*}
R_{1}=\left( 
\begin{array}{llll}
0 & 0 & 0 & 0 \\ 
0 & 0 & 0 & 0 \\ 
0 & 0 & 0 & 1 \\ 
0 & 0 & -1 & 0%
\end{array}%
\right) \;R_{2}=\left( 
\begin{array}{llll}
0 & 0 & 0 & 0 \\ 
0 & 0 & 0 & -1 \\ 
0 & 0 & 0 & 0 \\ 
0 & 1 & 0 & 0%
\end{array}%
\right) \;R_{3}=\left( 
\begin{array}{llll}
0 & 0 & 0 & 0 \\ 
0 & 0 & 1 & 0 \\ 
0 & -1 & 0 & 0 \\ 
0 & 0 & 0 & 0%
\end{array}%
\right)
\end{equation*}%
\begin{equation*}
L_{1}=\left( 
\begin{array}{llll}
0 & 1 & 0 & 0 \\ 
1 & 0 & 0 & 0 \\ 
0 & 0 & 0 & 0 \\ 
0 & 0 & 0 & 0%
\end{array}%
\right) \;L_{2}=\left( 
\begin{array}{llll}
0 & 0 & 1 & 0 \\ 
0 & 0 & 0 & 0 \\ 
1 & 0 & 0 & 0 \\ 
0 & 0 & 0 & 0%
\end{array}%
\right) \;L_{3}=\left( 
\begin{array}{llll}
0 & 0 & 0 & 1 \\ 
0 & 0 & 0 & 0 \\ 
0 & 0 & 0 & 0 \\ 
1 & 0 & 0 & 0%
\end{array}%
\right)
\end{equation*}%
\begin{equation*}
U_{1}^{\alpha }=\left( 
\begin{array}{llll}
0 & 0 & 0 & 0 \\ 
0 & 0 & 0 & 0 \\ 
0 & 0 & 0 & \alpha \\ 
0 & 0 & \alpha & 0%
\end{array}%
\right) \;U_{2}^{\alpha }=\left( 
\begin{array}{llll}
0 & 0 & 0 & 0 \\ 
0 & 0 & 0 & \alpha \\ 
0 & 0 & 0 & 0 \\ 
0 & \alpha & 0 & 0%
\end{array}%
\right) \;U_{3}^{\alpha }=\left( 
\begin{array}{llll}
0 & 0 & 0 & 0 \\ 
0 & 0 & \alpha & 0 \\ 
0 & \alpha & 0 & 0 \\ 
0 & 0 & 0 & 0%
\end{array}%
\right)
\end{equation*}%
\begin{equation*}
M_{1}^{\alpha }=\left( 
\begin{array}{llll}
0 & \alpha & 0 & 0 \\ 
-\alpha & 0 & 0 & 0 \\ 
0 & 0 & 0 & 0 \\ 
0 & 0 & 0 & 0%
\end{array}%
\right) \;M_{2}^{\alpha }=\left( 
\begin{array}{llll}
0 & 0 & \alpha & 0 \\ 
0 & 0 & 0 & 0 \\ 
-\alpha & 0 & 0 & 0 \\ 
0 & 0 & 0 & 0%
\end{array}%
\right) \;M_{3}^{\alpha }=\left( 
\begin{array}{llll}
0 & 0 & 0 & \alpha \\ 
0 & 0 & 0 & 0 \\ 
0 & 0 & 0 & 0 \\ 
-\alpha & 0 & 0 & 0%
\end{array}%
\right)
\end{equation*}%
\begin{equation*}
C_{1}^{\alpha }=\left( 
\begin{array}{llll}
0 & 0 & 0 & 0 \\ 
0 & \alpha & 0 & 0 \\ 
0 & 0 & 0 & 0 \\ 
0 & 0 & 0 & 0%
\end{array}%
\right) \;C_{2}^{\alpha }=\left( 
\begin{array}{llll}
0 & 0 & 0 & 0 \\ 
0 & 0 & 0 & 0 \\ 
0 & 0 & \alpha & 0 \\ 
0 & 0 & 0 & 0%
\end{array}%
\right) \;C_{3}^{\alpha }=\left( 
\begin{array}{llll}
0 & 0 & 0 & 0 \\ 
0 & 0 & 0 & 0 \\ 
0 & 0 & 0 & 0 \\ 
0 & 0 & 0 & \alpha%
\end{array}%
\right)
\end{equation*}%
\begin{equation*}
C_{0}^{\alpha }=\left( 
\begin{array}{llll}
\alpha & 0 & 0 & 0 \\ 
0 & 0 & 0 & 0 \\ 
0 & 0 & 0 & 0 \\ 
0 & 0 & 0 & 0%
\end{array}%
\right)
\end{equation*}%
$R_{i},L_{i}$ $(i=1,2,3)$ are generators of real rotations and real boosts.
For the other generators:

(1) $\alpha =i$ in the complex case;

(2) $\alpha =i,j,k$ with $i^{2}=j^{2}=k^{2}=-1$, $ij=-ji$, $ik=-ki$, $jk=-kj$%
, $ij=k$ and cyclic permutations in the quaternionic case;

(3) $\alpha =e_{1},e_{2},e_{3},e_{4},e_{5},e_{6},e_{7}$ with $e_{i}^{2}=-1$
and multiplication table

\begin{equation*}
\begin{tabular}{|c|c|c|c|c|c|c|c|}
\hline
& $e_{1}$ & $e_{2}$ & $e_{3}$ & $e_{4}$ & $e_{5}$ & $e_{6}$ & $e_{7}$ \\ 
\hline
$e_{1}$ & $-1$ & $e_{3}$ & $-e_{2}$ & $e_{5}$ & $-e_{4}$ & $-e_{7}$ & $e_{6}$
\\ \hline
$e_{2}$ & $-e_{3}$ & $-1$ & $e_{1}$ & $e_{6}$ & $e_{7}$ & $-e_{4}$ & $-e_{5}$
\\ \hline
$e_{3}$ & $e_{2}$ & $-e_{1}$ & $-1$ & $e_{7}$ & $-e_{6}$ & $e_{5}$ & $-e_{4}$
\\ \hline
$e_{4}$ & $-e_{5}$ & $-e_{6}$ & $-e_{7}$ & $-1$ & $e_{1}$ & $e_{2}$ & $e_{3}$
\\ \hline
$e_{5}$ & $e_{4}$ & $-e_{7}$ & $e_{6}$ & $-e_{1}$ & $-1$ & $-e_{3}$ & $e_{2}$
\\ \hline
$e_{6}$ & $e_{7}$ & $e_{4}$ & $-e_{5}$ & $-e_{2}$ & $e_{3}$ & $-1$ & $-e_{1}$
\\ \hline
$e_{7}$ & $-e_{6}$ & $e_{5}$ & $e_{4}$ & $-e_{3}$ & $-e_{2}$ & $e_{1}$ & $-1$
\\ \hline
\end{tabular}%
\ 
\end{equation*}%
for the octonionic case.

\begin{figure}[htb]
\centering
\includegraphics[width=0.5\textwidth]{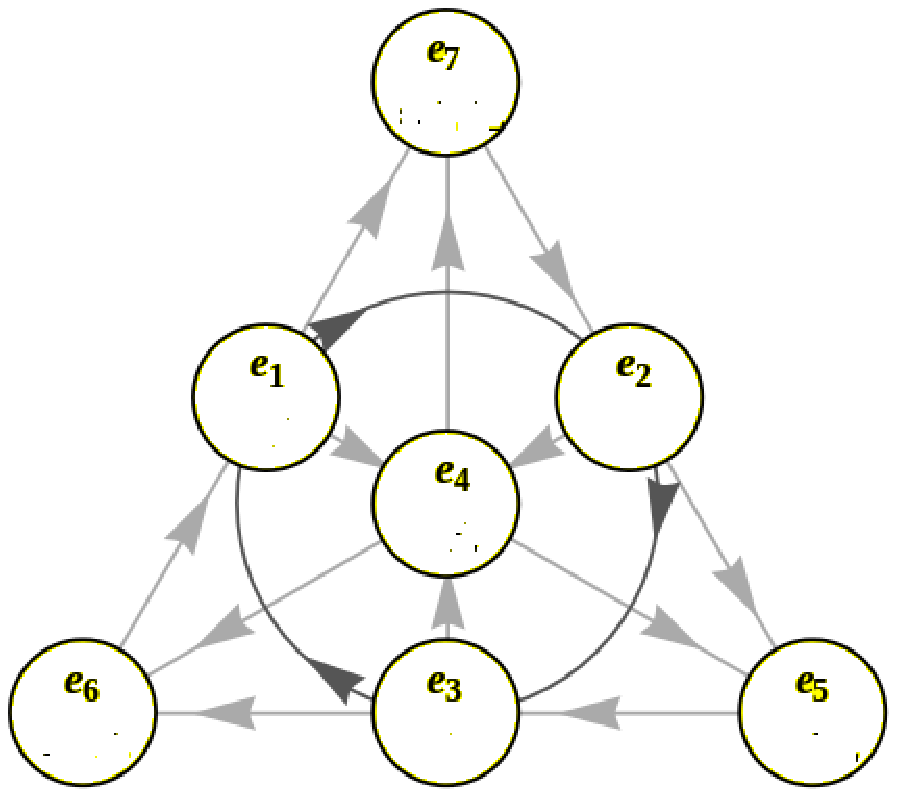}
\caption{}
\label{}
\end{figure}

The correspondence of
these generators to the $\left( M_{\mu \nu },N_{\mu \nu }\right) $ ones are:%
\begin{equation*}
R_{i}=\frac{1}{2}\epsilon
_{ijk}M_{jk};\;L_{i}=M_{0i};\;M_{i}=N_{0i};\;U_{i}=N_{jk};\;C_{\mu }=-\frac{1%
}{2}N_{\mu \mu }g_{\mu \mu }
\end{equation*}%
Other conventions are found in the literature for the labelling of the
octonionic imaginary units (see for example \cite{Baez-octo}). The above one
is used here because it is probably the one that better allows for a unified
treatment of the four normed division algebras. Notice also that, because of
consistency with the cyclic permutation order, the numeric labelling of the
generators of the complex Lorentz group differs from the one in \cite{Barut}.

\section*{\textbf{APPENDIX B}}

\textbf{Commutation table for the generators of the little (quasi)group in
the massive case }$\left( M_{0}^{2}>0\right) $

\begin{landscape}
\begin{tabular}{|l|l|l|l|l|l|l|l|l|l|}
\hline
& ${\small R}_{1}$ & ${\small R}_{2}$ & ${\small R}_{3}$ & ${\small U}%
_{1}^{\beta }$ & ${\small U}_{2}^{\beta }$ & ${\small U}_{3}^{\beta }$ & $%
{\small C}_{1}^{\beta }$ & ${\small C}_{2}^{\beta }$ & ${\small C}%
_{3}^{\beta }$ \\ \hline
${\small R}_{1}$ & ${\small 0}$ & ${\small -R}_{3}$ & ${\small R}_{2}$ & $%
{\small 2}\left( C_{2}^{\beta }-C_{3}^{\beta }\right) $ & ${\small U}%
_{3}^{\beta }$ & ${\small -U}_{2}^{\beta }$ & ${\small 0}$ & ${\small -U}%
_{1}^{\beta }$ & ${\small U}_{1}^{\beta }$ \\ \hline
${\small R}_{2}$ &  & ${\small 0}$ & ${\small -R}_{1}$ & ${\small -U}%
_{3}^{\beta }$ & ${\small 2}\left( -C_{1}^{\beta }+C_{3}^{\beta }\right) $ & 
${\small U}_{1}^{\beta }$ & ${\small U}_{2}^{\beta }$ & ${\small 0}$ & $%
{\small -U}_{2}^{\beta }$ \\ \hline
${\small R}_{3}$ &  &  & ${\small 0}$ & ${\small U}_{2}^{\beta }$ & ${\small %
-U}_{1}^{\beta }$ & ${\small 2}\left( C_{1}^{\beta }-C_{2}^{\beta }\right) $
& ${\small -U}_{3}^{\beta }$ & ${\small U}_{3}^{\beta }$ & ${\small 0}$ \\ 
\hline
${\small U}_{1}^{\alpha }$ &  &  &  & $%
\begin{array}{c}
{\small 0}\left( \alpha =\beta \right)  \\ 
\left( C_{2}+C_{3}\right) ^{\left[ \alpha ,\beta \right] } \\ 
\left( \alpha \neq \beta \right) 
\end{array}%
$ & $%
\begin{array}{c}
{\small R}_{3}\left( \alpha =\beta \right)  \\ 
{\small U}_{3}^{\left( \alpha \beta \right) }\left( \alpha \neq \beta
\right) 
\end{array}%
$ & $%
\begin{array}{c}
{\small -R}_{2}\left( \alpha =\beta \right)  \\ 
{\small U}_{2}^{\left( \alpha \beta \right) }\left( \alpha \neq \beta
\right) 
\end{array}%
$ & ${\small 0}$ & $%
\begin{array}{c}
{\small R}_{1}\left( \alpha =\beta \right)  \\ 
{\small U}_{1}^{\left( \alpha \beta \right) }\left( \alpha \neq \beta
\right) 
\end{array}%
$ & $%
\begin{array}{c}
{\small -R}_{1}\left( \alpha =\beta \right)  \\ 
{\small U}_{1}^{\left( \alpha \beta \right) }\left( \alpha \neq \beta
\right) 
\end{array}%
$ \\ \hline
${\small U}_{2}^{\alpha }$ &  &  &  &  & $%
\begin{array}{c}
{\small 0}\left( \alpha =\beta \right)  \\ 
\left( C_{1}+C_{3}\right) ^{\left[ \alpha ,\beta \right] } \\ 
\left( \alpha \neq \beta \right) 
\end{array}%
$ & $%
\begin{array}{c}
{\small R}_{1}\left( \alpha =\beta \right)  \\ 
{\small U}_{1}^{\left( \alpha \beta \right) }\left( \alpha \neq \beta
\right) 
\end{array}%
$ & $%
\begin{array}{c}
{\small -R}_{2}\left( \alpha =\beta \right)  \\ 
{\small U}_{2}^{\left( \alpha \beta \right) }\left( \alpha \neq \beta
\right) 
\end{array}%
$ & ${\small 0}$ & $%
\begin{array}{c}
{\small R}_{2}\left( \alpha =\beta \right)  \\ 
{\small U}_{2}^{\left( \alpha \beta \right) }\left( \alpha \neq \beta
\right) 
\end{array}%
$ \\ \hline
${\small U}_{3}^{\alpha }$ &  &  &  &  &  & $%
\begin{array}{c}
{\small 0}\left( \alpha =\beta \right)  \\ 
\left( C_{1}+C_{2}\right) ^{\left[ \alpha ,\beta \right] } \\ 
\left( \alpha \neq \beta \right) 
\end{array}%
$ & $%
\begin{array}{c}
{\small R}_{3}\left( \alpha =\beta \right)  \\ 
{\small U}_{3}^{\left( \alpha \beta \right) }\left( \alpha \neq \beta
\right) 
\end{array}%
$ & $%
\begin{array}{c}
{\small -R}_{3}\left( \alpha =\beta \right)  \\ 
{\small U}_{3}^{\left( \alpha \beta \right) }\left( \alpha \neq \beta
\right) 
\end{array}%
$ & ${\small 0}$ \\ \hline
${\small C}_{1}^{\alpha }$ &  &  &  &  &  &  & ${\small C}_{1}^{\left[
\alpha ,\beta \right] }$ & ${\small 0}$ & ${\small 0}$ \\ \hline
${\small C}_{2}^{\alpha }$ &  &  &  &  &  &  &  & ${\small C}_{2}^{\left[
\alpha ,\beta \right] }$ & ${\small 0}$ \\ \hline
${\small C}_{3}^{\alpha }$ &  &  &  &  &  &  &  &  & ${\small C}_{3}^{\left[
\alpha ,\beta \right] }$ \\ \hline
\end{tabular}
\end{landscape}
Here and in Appendix C, $X^{\left[ \alpha ,\beta \right] }$ means $%
2X^{\left\vert \alpha \beta \right\vert }sign\left( \alpha \beta \right) $
and $X^{\left( \alpha \beta \right) }$ means $X^{\left\vert \alpha \beta
\right\vert }sign\left( \alpha \beta \right) $.

\section*{\textbf{APPENDIX C}}

\textbf{Commutation table for the generators of the little (quasi)group in
the massless case }$\left( M_{0}^{2}=0,p_{\mu }\neq 0\right) $

\begin{eqnarray*}
l_{1} &=&L_{1}+R_{2};l_{2}=L_{2}-R_{1} \\
m_{1}^{\alpha } &=&M_{1}^{\alpha }+U_{2}^{\alpha };m_{2}^{\alpha
}=M_{2}^{\alpha }+U_{1}^{\alpha };m_{3}^{\alpha }=M_{3}^{\alpha
}+C_{3}^{\alpha }-C_{0}^{\alpha }
\end{eqnarray*}

\begin{tabular}{|l|l|l|l|l|l|l|l|l|l|}
\hline
& $l_{1}$ & $l_{2}$ & $m_{1}^{\beta }$ & $m_{2}^{\beta }$ & $m_{3}^{\beta }$
& $R_{3}$ & $U_{3}^{\beta }$ & $C_{1}^{\beta }$ & $C_{2}^{\beta }$ \\ \hline
$l_{1}$ & $0$ & $0$ & $2m_{3}^{\beta }$ & $0$ & $0$ & $l_{2}$ & $%
m_{2}^{\beta }$ & $m_{1}^{\beta }$ & $0$ \\ \hline
$l_{2}$ & $0$ & $0$ & $0$ & $2m_{3}^{\beta }$ & $0$ & $-l_{1}$ & $%
m_{1}^{\beta }$ & $0$ & $m_{2}^{\beta }$ \\ \hline
$m_{1}^{\alpha }$ &  &  & $m_{3}^{\left[ \alpha ,\beta \right] }$ & $0$ & $0$
& $m_{2}^{\alpha }$ & $%
\begin{array}{c}
-l_{2}\left( \alpha =\beta \right) \\ 
m_{2}^{\left( \alpha \beta \right) }\left( \alpha \neq \beta \right)%
\end{array}%
$ & $%
\begin{array}{c}
-l_{1}\left( \alpha =\beta \right) \\ 
m_{1}^{\left( \alpha \beta \right) }\left( \alpha \neq \beta \right)%
\end{array}%
$ & $0$ \\ \hline
$m_{2}^{\alpha }$ &  &  &  & $m_{3}^{\left[ \alpha ,\beta \right] }$ & $0$ & 
$-m_{1}^{\alpha }$ & $%
\begin{array}{c}
-l_{1}\left( \alpha =\beta \right) \\ 
m_{1}^{\left( \alpha \beta \right) }\left( \alpha \neq \beta \right)%
\end{array}%
$ & $0$ & $%
\begin{array}{c}
-l_{2}\left( \alpha =\beta \right) \\ 
m_{2}^{\left( \alpha \beta \right) }\left( \alpha \neq \beta \right)%
\end{array}%
$ \\ \hline
$m_{3}^{\alpha }$ &  &  &  &  & $0$ & $0$ & $0$ & $0$ & $0$ \\ \hline
$R_{3}$ &  &  &  &  &  & $0$ & $2\left( C_{1}^{\beta }-C_{2}^{\beta }\right) 
$ & $-u_{3}^{\beta }$ & $u_{3}^{\beta }$ \\ \hline
$U_{3}^{\alpha }$ &  &  &  &  &  &  & $\left( C_{1}+C_{2}\right) ^{\left[
\alpha ,\beta \right] }$ & $%
\begin{array}{c}
r_{3}\left( \alpha =\beta \right) \\ 
u_{3}^{\left( \alpha \beta \right) }\left( \alpha \neq \beta \right)%
\end{array}%
$ & $%
\begin{array}{c}
-r_{3}\left( \alpha =\beta \right) \\ 
u_{3}^{\left( \alpha \beta \right) }\left( \alpha \neq \beta \right)%
\end{array}%
$ \\ \hline
$C_{1}^{\alpha }$ &  &  &  &  &  &  &  & $C_{1}^{\left[ \alpha ,\beta \right]
}$ & $0$ \\ \hline
$C_{2}^{\alpha }$ &  &  &  &  &  &  &  &  & $C_{2}^{\left[ \alpha ,\beta %
\right] }$ \\ \hline
\end{tabular}

\section*{Appendix D: Half-integer spin states in homogeneous spaces}

Half-integer spin states (of the subgroup generated by $\left\{
R_{i}\right\} $) are not contained in the irreducible representations of $%
U\left( 3\right) $ (the little group of massive states of $U\left(
3,1\right) $), nevertheless half-spin state representations may be
associated to $U\left( 3\right) $ in a nonlinear way. Here one considers the
complex Poincar\'{e} group case.

The set $\left\{ R_{i}\right\} $ generates a $SU\left( 2\right) $ subgroup
of $U\left( 3\right) $. With a coset decomposition of $U\left( 3\right) $
one obtains a six-dimensional homogeneous space $M=U\left( 3\right)
/SU\left( 2\right) $. Let us label the cosets by the letter $p$ and for each
coset chose an element $\sigma \left( p\right) \in U\left( 3\right) $ such
that a generic element of that coset is $\sigma \left( p\right) h$ with $%
h\in SU\left( 2\right) $. To the homogeneous space $M$ associate a vector
bundle $\Gamma $ with base $M$ and fibers $V$ carrying an half-integer
representation of $SU\left( 2\right) $. Locally the bundle is $N\left(
p\right) \times V$, $N\left( p\right) $ being a\ neighborood of $p\in M$. An
arbitrary element of $\Gamma $ is $\Phi \left( p,\alpha \right) $ where $%
\alpha $ carries the quantum numbers of the $SU\left( 2\right) $
representation.

The action of an arbitrary element of $g\in U\left( 3\right) $ on $\Phi
\left( p,\alpha \right) $ is obtained as follows. Notice that%
\begin{equation*}
g\sigma \left( p\right) =\sigma \left( p^{\prime }\right) h\left(
g,p,p^{\prime }\right)
\end{equation*}%
with $h\left( g,p,p^{\prime }\right) \in SU\left( 2\right) $. Then%
\begin{equation*}
g\Phi \left( p,\alpha \right) =D\left( h\left( g,p,p^{\prime }\right)
\right) \circ \Phi \left( p^{\prime },\alpha \right)
\end{equation*}%
$D\left( .\right) $ being a representation matrix of $SU\left( 2\right) $.
If instead of the states $\Phi $ one cone considers sections $\psi \left(
p\right) \in V$ of the bundle%
\begin{equation*}
g\circ \psi \left( p\right) =D\left( h^{-1}\left( g,p,p^{\prime }\right)
\right) \circ \psi \left( \tau _{g}^{-1}p^{\prime }\right)
\end{equation*}%
$\tau _{g}:p\rightarrow p^{\prime }$ being the action of $g$ on $M$.

This construction implements a representation of $U\left( 3\right) $
carrying half-integer spin states of the $SU\left( 2\right) $ generated by $%
\left\{ R_{i}\right\} $. However, this representation cannot be reduced into
irreducible representations of $U\left( 3\right) $.

The physical interpretation is that, when operating on the half-integer spin
states of the real slice Poincar\'{e} group, the additional generators of
the complex Lorentz group instead of carrying the states to a different real
slice, move them in an internal space of dimension six. In this sense the
kinematical transformations of the complex Lorentz group become internal
symmetries. Notice however that this situation is mathematically different
from, for example, assuming an internal $SU\left( 3\right) $ colour symmetry
and generating a colour triplet and anti-triplet space. Here the label space
(the base of the bundle) is six-dimensional but the representation is in
fact infinite-dimensional.

For the quaternionic and octonionic space-times the situation is similar in
the sense that also half-integer spin states do not appear as elementary
states, but the dimension of the internal spaces (the base of the bundles)
is larger.

\end{document}